\documentclass[aps,twocolumn,prl,floatfix,square,comma,numbers,showpacs]{revtex4}
\usepackage{amssymb}
\usepackage[latin9]{inputenc}
\usepackage{amsmath}
\usepackage{amscd}
\usepackage{graphicx}
\usepackage{subfigure}
\usepackage{float}

\begin{document}

\title{Quantum delayed-choice experiment with a beam splitter in a quantum
superposition}
\author{Shi-Biao Zheng$^1$}
\email{t96034@fzu.edu.cn}
\author{You-Peng Zhong$^2$, Kai Xu$^2$, Qi-Jue Wang$^2$, H. Wang$^2$,
Li-Tuo~Shen$^1$, Chui-Ping Yang$^3$}
\author{John M. Martinis$^4$}
\email{martinis@physics.ucsb.edu}
\author{A. N. Cleland$^5$}
\email{anc@uchicago.edu}
\author{Si-Yuan Han$^{6,7}$}

\begin{abstract}
A quantum system can behave as a wave or as a particle, depending on the
experimental arrangement. When for example measuring a photon using a
Mach-Zehnder interferometer, the photon acts as a wave if the second
beam-splitter is inserted, but as a particle if this beam-splitter is
omitted. The decision of whether or not to insert this beam-splitter can be
made after the photon has entered the interferometer, as in Wheeler's famous
delayed-choice thought experiment. In recent quantum versions of this
experiment, this decision is controlled by a quantum ancilla, while the beam
splitter is itself still a classical object. Here we propose and realize a
variant of the quantum delayed-choice experiment. We configure a
superconducting quantum circuit as a Ramsey interferometer, where the
element that acts as the first beam-splitter can be put in a quantum
superposition of its active and inactive states, as verified by the negative
values of its Wigner function. We show that this enables the wave and
particle aspects of the system to be observed with a single setup, without
involving an ancilla that is not itself a part of the interferometer. We
also study the transition of this quantum beam-splitter from a quantum to a
classical object due to decoherence, as observed by monitoring the
interferometer output.
\end{abstract}

\affiliation{$^{1}$ Department of Physics, Fuzhou University, Fuzhou 350116,
China \\$^{2}$ Department of Physics, Zhejiang University, Hangzhou 310027,
China \\$^{3}$ Department of Physics, Hangzhou Normal University, Hangzhou
310036, China \\$^{4}$ Department of Physics, University of California,
Santa Barbara, California 93106, USA \\$^{5}$ Institute for Molecular
Engineering, University of Chicago, Chicago, Illinois 60637, USA\\$^{6}$
Department of Physics and Astronomy, University of Kansas, Lawrence, Kansas
66045, USA\\$^{7}$ Beijing National Laboratory for Condensed Matter Physics,
Institute of Physics, Chinese Academy of Sciences, Beijing 100190, China}

\maketitle






The wave-particle duality is one of the fundamental mysteries that lie at
the heart of quantum mechanics. However, these two incompatible aspects
cannot be observed simultaneously, as captured by Bohr's principle of
complementarity $[1-5]$: Particle-like versus wave-like outcomes are
selected by experimental arrangements that are mutually exclusive. This is
well illustrated by the Mach-Zehnder interferometer, as shown in Fig. 1(a).
Split by the first beam splitter (BS$_1$), a photon travels along two paths,
0 and 1. The relative phase $\theta $ between the quantum states associated
with these paths is tunable. In the presence of the second beam splitter (BS$%
_2$), the two paths are recombined and the probability for detecting the
photon in the detector D$_0$ or D$_1$ is a sinusoidal function of $\theta $,
exhibiting wave-like interference fringes. On the other hand, in the absence
of BS$_2$, the experiment reveals which path the photon followed, and the
photon is detected in one or the other detector with equal probability 1/2,
thus behaving as a particle.

One can argue that the behavior of the photon is pre-determined by the
experimental arrangement, where the presence or absence of the second beam
splitter affects the photon prior to its entering the interferometer. The
possibility of such a causal link is precluded in Wheeler's delayed-choice
experiment $[6-8]$, in which the observer randomly chooses whether to insert
BS$_2$, and thus whether to perform an interference or a which-path
experiment, after the photon has passed through BS$_1$. Therefore, the
photon could not ``know'' in advance which behavior it should exhibit.
Wheeler's delayed-choice experiment has been demonstrated previously $[9-13]$%
, where the space-like separation between the setup selection and the photon
entry into the interferometer was achieved in Ref. [12]. Recently, a quantum
version of the delayed-choice experiment was suggested $[14]$, in which the
action of BS$_2$ is controlled by a quantum ancilla. The scheme is
illustrated in Fig.~1(b), where each beam splitter is replaced by a Hadamard
operation H, and the second Hadamard operation is conditionally applied
following the phase shift $\theta $. The quantum system exhibits wave-like
behavior if the ancilla is in its $|1\rangle $ state and the second Hadamard
is applied; if the ancilla is instead in $|0\rangle $, the second Hadamard
is not performed, and the system displays particle-like behavior. When the
ancilla is prepared in a superposition of $\left| 0\right\rangle $ and $%
\left| 1\right\rangle $, the result is a superposition of wave-like and
particle-like states, entangled with the ancilla $[14]$. This clearly
precludes the system `knowing' in advance which setup will be selected; the
quantum superposition effectively replaces the need for Wheeler's space-like
separation of the photon entry into the interferometer and the measurement
selection. Instead, the wave-like and particle-like behaviors are
post-selected by measuring the ancilla in its $(|0\rangle ,|1\rangle )$
basis. This allows the complementary wave and particle behaviors to emerge
from a single experimental set-up, showing that the complementarity really
resides in the experimental data, rather than resulting from the
experiment's physical arrangement.

Quantum delayed-choice experiments have been performed in both NMR $[15,16]$
and optical systems $[17-19]$. In the optical experiments reported in Refs.
[18] and [19], the quantum correlation between the test photon and the
quantum ancilla was verified by the violation of a Bell inequality; we note
that these measurements are subject to the fair-sampling assumption, that
is, the large number of coincident photon pairs that fail to be registered
are assumed to obey the same statistics as those that are recorded.

\begin{figure}[tbp]
\includegraphics[width=0.65\columnwidth]{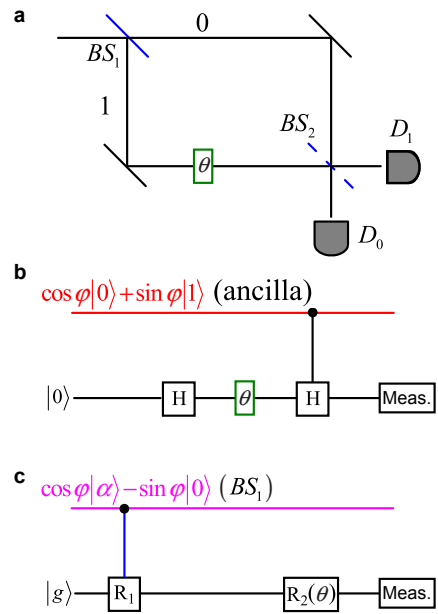}
\caption{(Color online) (a) Schematic of Wheeler's delayed-choice
thought experiment. A photon's state is transformed by the first
beam splitter BS$_1$, into a superposition of two paths, 0 and 1,
followed by a tunable phase shift $\theta $. While the photon is
inside the interferometer, the observer decides whether or not to
insert a second beam splitter BS$_2$, thereby delaying the decision
of whether to measure an interference pattern (wave aspect) or to
obtain which-path information (particle aspect), as revealed by the
probability of detecting the photon at detector D$_0$ and D$_1$ as a
function of $\theta $. (b) Schematic of the quantum delayed-choice
experiment using a controlling quantum ancilla. The effect of each
beam splitter is represented by a Hadamard operation H, with the
relative phase shift $\theta $ occurring in-between these two
operations. The second Hadamard is controlled by the ancilla, which
can be prepared in a superposition state, allowing simultaneous
preparation of the wave and particle aspects of the test system. (c)
Schematic of the delayed choice experiment, implemented with a
superconducting beam splitter that is put into a cat-like
superposition state. The test qubit, initially in the ground state
$\left| g\right\rangle $, is subjected to two operations, R$_1$ and
R$_2(\theta )$, equivalent in combination to the two Hadamard
operations plus the phase shift $\theta $. R$_1$ is implemented
(active) when the test qubit interacts with a superconducting
resonator occupied by a classical coherent field $|\alpha \rangle $.
R$_2(\theta )$ is implemented by a classical microwave pulse. The
final probability of measuring the qubit in the ground state $\left|
g\right\rangle $ or excited state $\left|
e\right\rangle $, for active R$_1$, then depends on the relative phase $%
\theta $, thus exhibiting Ramsey interference. When the
superconducting resonator is instead in the ground state $|0\rangle
$, R$_1$ is not implemented (inactive), and no $\theta $-dependent
Ramsey interference occurs. The wave and particle aspects of the
qubit can be superposed by preparing the resonator in a quantum
superposition of ''filled'' $|\alpha \rangle $ and ''empty''
$|0\rangle $ states, generating an output state that encodes both
particle- and wave-like behavior. }\end{figure}

Here we theoretically propose and experimentally implement a variant of the
quantum delayed-choice experiment, using a superconducting Ramsey
interferometer and the process shown in Fig.~1(c); we note that the Ramsey
and Mach-Zehnder interferometers are physically different, but actually
implement the same quantum circuit $[20]$. In our interferometer, a
two-level quantum system (a superconducting qubit), with ground state $%
\left| g\right\rangle $ and excited state $\left| e\right\rangle $, which
correspond to the paths 0 and 1 in the optical interferometer, is subjected
to two Ramsey rotations, R$_1$ and R$_2(\theta )$, which correspond to the
two beam-splitters of the Mach-Zehnder interferometer but are induced by
microwave fields. The tunable phase shift $\theta $ between the two
interfering paths is incorporated into the microwave field that produces R$%
_2(\theta )$. R$_1$ is generated by the microwave field stored in a
superconducting resonator. If the resonator is ``filled'' by exciting it
into the appropriate microwave coherent state $|\alpha \rangle $, the qubit
will undergo a Hadamard transformation by interacting with this field. This
rotates the input state from $\left| g\right\rangle $ into a superposition
of $\left| g\right\rangle $ and $\left| e\right\rangle $, producing two
paths in the Hilbert space, as with the first beam-splitter in the
Mach-Zehnder interferometer. These two paths are then recombined by R$%
_2(\theta )$, and that recombination results in $\theta $-dependent
wave-like interference fringes in the probabilities of $|g\rangle $ and $%
|e\rangle $. If the resonator is instead ``empty'', i.e. in its ground state
$|0\rangle $, no rotation R$_1$ occurs, and the qubit remains in $|g\rangle $%
. The second rotation R$_2(\theta )$ then produces a superposition of $%
|g\rangle $ and $|e\rangle $, but the probabilities of $|g\rangle $ and $%
|e\rangle $ have no $\theta $ dependence, corresponding to particle-like
behavior. We note that the required conditional dynamics cannot be realized
when the order of the two rotations in Fig. 1(c) is inverted: To realize the
resonator-state-dependent rotation, the qubit should be in the state $\left|
g\right\rangle $\ before interaction with the resonator, so that the
resonator's ground state $\left| 0\right\rangle $\ cannot induce any
rotation on the qubit; however, when the unconditional rotation R$_2(\theta
) $\ is applied first, before the qubit-resonator interaction the qubit has
a probability of 1/2 to be excited by R$_2(\theta )$\ to the state $\left|
e\right\rangle $, to which the resonator's ground state can produce a
rotation through the vacuum Rabi oscillation [24].

The wave and particle behavior of the qubit can be investigated
simultaneously by preparing the resonator in a superposition of its
``filled'' $|\alpha \rangle $ and ``empty'' $|0\rangle $ states, i.e. in a
Schr\"odinger cat state $[21]$. The nonclassical nature of this quantum beam
splitter (QBS) can be revealed by measuring the negative values of its
Wigner function (WF) $[22]$, which is the resonator's quasiprobability
distribution in phase space. This enables one to verify the existence of a
coherent quantum superposition of the filled and empty, or active and
inactive, states of the QBS, without performing a Bell measurement. In this
case, after R$_2$, the wave and particle states of the qubit are entangled
with the state of the resonator. This is in contrast to previous
realizations of quantum delayed-choice experiments $[15-19]$, where the
wave-particle response of the quantum system is determined by the state of a
quantum ancilla, which is not itself a part of the interferometer, but
instead determines the action (or inaction) of one of the two
beam-splitters. In these experiments, the test system is entangled with the
ancilla through the conditional action of the beam-splitter, while the beam
splitter remains a fully classical object. With our setup, we can also
investigate the transition of the QBS to a classical beam-splitter due to
its intrinsic environmentally-induced decoherence. As far as we know, this
is the first interference experiment in which one beam splitter in an
interferometer is in a coherent superposition of its active and inactive
states. We note that our setup, using a temporally-based Ramsey
interferometer, does not permit the space-like separation required for
Wheeler's original gedanken experiment.

The rotation R$_1$ of the qubit is produced by the microwave field stored in
the resonator, which is resonantly coupled to the qubit transition $\left|
g\right\rangle \leftrightarrow \left| e\right\rangle $ with coupling
strength $\Omega $. If the resonator is in the state $|\alpha \rangle $, a
coherent microwave photon state with amplitude $\alpha $ and mean photon
number $\langle n\rangle =|\alpha |^2$, the qubit exchanges energy with the
resonator and Rabi-oscillates between $\left| g\right\rangle $ and $\left|
e\right\rangle $. For simplicity, we assume $\alpha $ is real and positive.
When $\alpha \gg 1$, after an interaction time $t_\alpha =\pi /(4\alpha
\Omega )$, the qubit state is approximately $|\psi _\alpha \rangle =(\left|
g\right\rangle -i\left| e\right\rangle )/\sqrt{2}$, with the resonator field
left close to its original state $\left| \alpha \right\rangle $. Numerical
simulation shows that the approximation is good even for moderate values of $%
\alpha $. This result has a simple qualitative explanation: When the
resonator's coherent field is not too weak, its Poissonian photon-number
distribution and hence its state is insensitive to a one-photon change. If
the resonator is instead in its ground state $|0\rangle $, the rotation R$_1$
does not occur and the qubit remains in $|g\rangle $. The second pulse R$%
_2(\theta )$, generated by a classical microwave pulse, performs the
transformations $\left| g\right\rangle \rightarrow (\left| g\right\rangle
-ie^{-i\theta }\left| e\right\rangle )/\sqrt{2}$\ and $\left| e\right\rangle
\rightarrow (\left| e\right\rangle -ie^{i\theta }\left| g\right\rangle )/%
\sqrt{2}$. When acting on the state $|\psi _\alpha \rangle $, this results
in the state
\begin{equation}
|\psi _w\rangle =-i\left[ \sin (\theta /2)e^{i\theta /2}|g\rangle +\cos
(\theta /2)e^{-i\theta /2}|e\rangle \right] .
\end{equation}
The probability of measuring the qubit in $|e\rangle $ is then $\cos
^2(\theta /2)$, showing the $\theta $-dependent interference associated with
wave-like behavior. If R$_1$ did not occur, due to the microwave resonator
being in its ground state $|0\rangle $, the final state after the second
rotation R$_2$ is
\begin{equation}
\left| \psi _p\right\rangle =(|g\rangle -ie^{-i\theta }|e\rangle )/\sqrt{2}.
\end{equation}
The probability to be in $|g\rangle $ or $|e\rangle $ is 1/2, representing
particle-like behavior without the $\theta $-dependent interference effects.

Now we suppose that the resonator is instead initially prepared in a cat
state
\begin{equation}
\left| \psi _{b,i}\right\rangle =\mathcal{N}(\cos \varphi \left| \alpha
\right\rangle -\sin \varphi \left| 0\right\rangle ),
\end{equation}
where $\mathcal{N}=\left[ 1-e^{-\left| \alpha \right| ^2/2}\sin (2\varphi
)\right] ^{-1/2}$. After the two rotations R$_1$ and R$_2(\theta )$, the
qubit-QBS system will then approximately be in the entangled state
\begin{equation}
\left| \psi _{q+b,f}\right\rangle \simeq \mathcal{N}_t(\cos \varphi \left|
\psi _w\right\rangle \left| \alpha \right\rangle -\sin \varphi \left| \psi
_p\right\rangle \left| 0\right\rangle ),
\end{equation}
where $\mathcal{N}_t=\left[ 1-e^{-\left| \alpha \right| ^2/2}\sin (2\varphi
)/\sqrt{2}\right] ^{-1/2}$. The probability $P_e$ for finding the qubit in
the state $\left| e\right\rangle $ is then close to
\begin{equation}
P_e\simeq \mathcal{N}_t^2\left[ \frac 12\sin ^2\varphi +\cos ^2\varphi \cos
^2\frac \theta 2\right] .
\end{equation}
The Ramsey interference pattern thus simultaneously exhibits the wave and
particle behaviors, through the presence and absence, respectively, of $%
\theta $ dependence in the two terms that make up $P_e$. The quantum
coherence between the two state components $\left| \alpha \right\rangle $
and $\left| 0\right\rangle $ precludes the possibility of the qubit
``knowing'' in advance what type of experiment will be performed. We note
that for $\varphi =0$ this interferometer is equivalent to the cavity QED
analogue described in Ref. [5] except there the qubit is initially in $%
\left| e\right\rangle $.

The detailed implementation of this quantum delayed-choice experiment
involves a superconducting resonator and two tunable superconducting phase
qubits, one qubit serving as the test qubit and the second control qubit
used to program the state of the superconducting resonator, as well as to
read out the resonator at the end of each experiment. The two qubits are
coupled to the resonator with on-resonance coupling strengths $\Omega $ and $%
\Omega ^{\prime }$, respectively, and the effective coupling of each qubit
to the resonator can be effectively switched on or off by tuning the qubit
on- or off-resonance with the resonator. The experimental apparatus is
identical to that described in Ref. [23]. Using this apparatus, we can
arrive at the state $\left| \psi _{b,i}\right\rangle $ of Eq. (3) with $%
\alpha =2$, and arbitrary values of $\varphi $ (Ref. [24]; see Supplementary
Material). For $\varphi =\pi /4$, the fidelity $F=\left\langle \psi
_{b,i}\right| \rho _{b,i}\left| \psi _{b,i}\right\rangle =0.726\pm 0.028$ ($%
\rho _{b,i}$ is the density operator corresponding to the output state). It
would be preferable to obtain states with larger photon amplitude $\alpha $,
but we are limited by the small nonlinearity of our phase qubits to $\alpha
\lesssim 2$. Fortunately, numerical simulations show that the overlap
between the final qubit-QBS state and $\left| \psi _{q+b,f}\right\rangle $
of Eq. (4) is higher than $0.98$ even for this value of $\alpha $ (ignoring
experimental imperfections). After generating $\left| \psi
_{b,i}\right\rangle $, we resonantly couple the test qubit to the resonator
for a time $t_\alpha $, and then apply a carefully tuned, on-resonance
microwave $\pi /2$\ pulse to the qubit, thus implementing the rotation R$%
_2(\theta )$. Figure 2(a) displays the measured probability $P_e$ as a
function of $\theta $ and $\varphi $, which clearly demonstrates the
morphing between particle and wave behavior.
\begin{figure}[tbp]
\includegraphics[width=1\columnwidth]{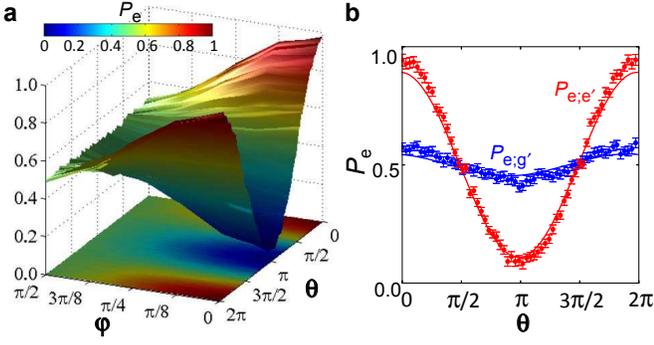}
\caption{(Color online) Measured Ramsey interference signal. (a) Probability $%
P_e $, defined in Eq. (5), versus $\theta $ and $\varphi $. The
amplitude of
the coherent state component of the cat state $|\psi _{b,i}\rangle $ is $%
\alpha =2$. The transition between the wave and particle behaviors is
clearly apparent. (b) Probabilities $P_{e;e^{\prime }}$ and $P_{e;g^{\prime
}}$, versus $\theta $, for detecting the test qubit in $|e\rangle $ given
that the control qubit is detected in $|e^{\prime }\rangle $ and $|g^{\prime
}\rangle $, respectively. The control qubit's state reflects the QBS state,
as this qubit is measured after it has interacted with the post-R$_2$
resonator state for a time $\pi /(2\alpha \Omega ^{\prime })$. Data are for $%
\varphi =\pi /4$. Error bars indicate the statistical variance.
Lines are simulations taking into account imperfections in preparing
$|\psi _{b,i}\rangle $.}\end{figure}
To more precisely understand
this behavior, we need to examine the state of the QBS. This is
achieved using the control qubit, initially in the state $|g^{\prime
}\rangle $, and tuning it into resonance with the resonator for a
time $\pi /\left( 2\alpha \Omega ^{\prime }\right) $. If the
resonator contains a coherent field $|\alpha \rangle $, the control
qubit will absorb a photon and undergo the transition $|g^{\prime
}\rangle \rightarrow |e^{\prime }\rangle $, while if the resonator
is in $|0\rangle $, the qubit remains in $|g^{\prime }\rangle $. The
test qubit behavior is post-selected by correlating its measurements
with the outcomes of the control qubit measurements. We note that
$\left|
\alpha \right\rangle $ is not strictly orthogonal to $\left| 0\right\rangle $%
, with the overlap $\left| \left\langle \alpha \right| \left. 0\right\rangle
\right| ^2\approx 10^{-2}$ for $\alpha =2$, so that these two components
cannot be unambiguously discriminated; there is a small probability that the
detection of $\left| 0\right\rangle $ actually comes from $\left| \alpha
\right\rangle $. With $\alpha $ somewhat larger, this overlap will become
negligible: For $\alpha =3$, this probability is only about $10^{-4}$.

Figure~2(b) shows the measured probabilities $P_{e;e^{\prime }}$ and $%
P_{e;g^{\prime }}$ for detecting the test qubit in the state $|e\rangle $
conditioned upon the detection of the control qubit in $|e^{\prime }\rangle $
and $|g^{\prime }\rangle $, respectively; these are measured as a function
of $\theta $. Here the parameter $\varphi $ is $\pi /4$, corresponding to
the QBS being in an equal superposition of its active and inactive states.
As expected, $P_{e;e^{\prime }}$ exhibits Ramsey interference fringes, with
the contrast reaching $0.839$, while $P_{e;g^{\prime }}$ remains almost
constant. The slight oscillations in $P_{e;g^{\prime }}$\ are mainly due to
the fact that the amplitude of the coherent state component $|\alpha \rangle
$\ is somewhat small, so that the control qubit has a small probability of
remaining in the ground state after the interaction even when the resonator
is in $|\alpha \rangle $. On one hand, the coherent state contains a vacuum
state component, which is decoupled from the qubit state $\left| g^{^{\prime
}}\right\rangle $. On the other hand, not all the other superposed Fock
states can make the control qubit flip to the excited state $\left|
e^{^{\prime }}\right\rangle $\ with a unity probability after the
interaction, owing to the photon-number dependence of the Rabi frequency.
These two reasons account for the $\theta $-dependent interference effect in
$P_{e;g^{^{\prime }}}$, with a fringe contrast measured to be 0.19. It is
noted that taking into account the imperfect initial state as we prepared
experimentally, we could verify that the measured fringe data (dots with
error bars) are in good agreement with the numerical simulation (lines) in
Fig. 2(b). With an increase in the coherent state amplitude $\alpha $, the
state $\left| \alpha \right\rangle $\ is more clearly distinguished from $%
\left| 0\right\rangle $, and as a result the unwanted oscillations decay
dramatically: For $\alpha =3$, the calculated fringe contrast is reduced to
0.017.\emph{\ }
\begin{figure}[tbp]
\includegraphics[width=1\columnwidth]{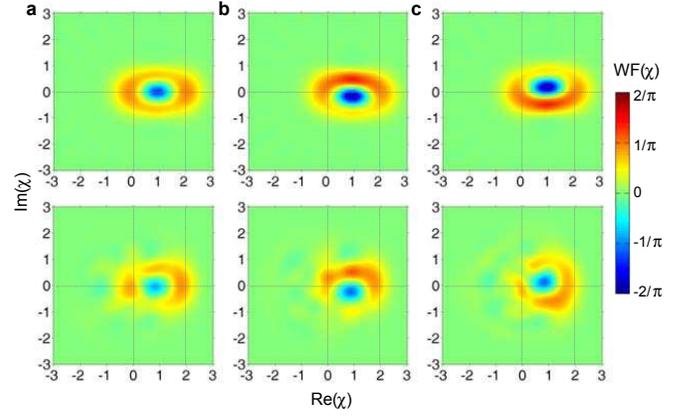}
\caption{(Color online) Wigner tomography of the quantum beam splitter. The
parameters are $\alpha =2$, $\theta =\pi /2$ and $\varphi =\pi /4$. The
quantum beam splitter Wigner functions (WFs) are displayed for three cases:
(a) Test qubit state is not read out; (b) Test qubit is measured in $%
|g\rangle $; (c) Test qubit is measured in $|e\rangle $. The simulated and
measured WFs are shown in the upper and lower rows, respectively, as a
function of the coordinate $\chi $ in resonator phase space. Experimental
imperfections are not included in the numerical simulations. The minimum
values of the three measured WFs are $-0.258\pm 0.030$, $-0.342\pm 0.027$,
and $-0.336\pm 0.028$, respectively. }
\label{Fig.3}
\end{figure}

It should be noted that the same statistical data can be produced if the
resonator field is in the classical mixture $\cos ^2\varphi \left| \alpha
\right\rangle \left\langle \alpha \right| +\sin ^2\varphi \left|
0\right\rangle \left\langle 0\right| $. To exclude the classical
interpretation, the existence of quantum coherence between $\left| \alpha
\right\rangle $ and $\left| 0\right\rangle $ should be verified. The quantum
state of the resonator field can be characterized by measuring its Wigner
function (WF) $W(\chi )$, which describes the quasiprobability distribution
of the microwave field in resonator phase space $[22]$. The WF associated
with the density operator $\rho _b$ is defined as
\begin{equation}
W(\chi )=\frac 2\pi Tr\left[ e^{i\pi a^{\dagger }a}D(-\chi )\rho _bD(\chi
)\right] ,
\end{equation}
where $\chi $ is the (complex) coordinate in phase space, and $D$ is the
displacement operator. This quantity is always non-negative for the QBS in a
classical mixture; the observation of negative values in regions of phase
space is a signature of quantum interference. The Wigner function of the QBS
was measured using the control qubit, following the procedure developed in
Ref. [24]. In Fig.~3(a), we display the WF after performing the second
rotation R$_2(\theta )$, but without reading out the state of the test
qubit, while in Figs.~3(b) and (c) we show the WFs measured with the test
qubit having been measured in $|g\rangle $ and $|e\rangle $, respectively,
all for $\theta =\pi /2$ and $\varphi =\pi /4$. Simulated and measured WFs
are shown in the upper and lower panels, respectively. Experimental
imperfections are not included in the numerical simulations. Since the qubit
wave state $|\psi _w\rangle $ is not orthogonal to the particle state $|\psi
_p\rangle $, there is some quantum coherence between $|\alpha \rangle $ and $%
|0\rangle $, even when the test qubit is traced out. As a result, the WF
exhibits a strongly nonclassical feature around $\chi =1$, as shown in
Fig.~3(a). The measured WF has a minimum value of $-0.258\pm 0.030$ at $\chi
=0.84-0.03i$. In Fig.~3(a), the shapes of the calculated and measured WFs
agree well, demonstrating that the measured negative quasi-probabilities are
due to quantum interference between $|\alpha \rangle $ and $|0\rangle $. The
existence of quantum coherence between these two state components implies
that their correlation with the behavior of the test qubit shown in Fig.
2(b) is nonclassical. Without reading out the test qubit's state, the WF is
independent of the value of $\theta $, since any local unitary operation on
the test qubit does not affect the QBS state\emph{\ }after their
interaction. When the test qubit's state is measured, the minimum value of
the WF becomes more negative, implying the enhancement of quantum
interference between the active and inactive states of the QBS. We have also
measured the corresponding WFs for $\varphi =\pi /8$ and $3\pi /8$ (see Fig.
S.2 in the Supplementary Material for details). As expected, for both cases
there exists a region where the WF has negative values, and the quantum
interference is enhanced when the test qubit's state is measured. These
results further show that there exists entanglement between the quantum
beam-splitter and the test qubit. We note the final QBS-qubit state was
realized deterministically and the qubit measurement was done in a
single-shot manner. Another benefit of this experimental implementation is
that it allows the observation of the transition from a quantum to a
classical beam splitter, as shown in the Supplementary Material.

We have proposed and carried out a quantum delayed-choice experiment for a
qubit interacting with a Ramsey interferometer, achieved by preparing one of
the two Ramsey beam splitters in a superposition of its active and inactive
states. Unlike previous experiments, here the beam splitter is really a
quantum object, and the qubit behavior is clearly correlated with the
quantum state of this beam splitter, significantly different from situations
where an ancilla controls the transformation produced by a classical beam
splitter. The quantum nature of the QBS is unambiguously verified by the
negative values of its Wigner function. We have also observed variations in
the Ramsey fringe contrast and the corresponding negativity of the Wigner
function as a function of delay, illustrating the quick damping of the
quantum coherence of the QBS. Using qubits with stronger nonlinearity and
longer coherence times, we plan to increase the size and fidelity of the cat
state and to explore the gradual transition from a quantum to a classical
measuring device. This experiment was realized using a circuit QED system;
however, similar experiments could be performed using microwave cavity QED $%
[25,26]$ or ion-trap setups $[27,28]$. We further note that the idea of
producing a conditional rotation on a qubit with an oscillator in a
mesoscopic superposition may be useful for generation of important entangled
states. This conditional dynamics, together with the qubit-oscillator
quantum state transfer $[29]$, could be used to produce entanglement between
two oscillators. Another example is the generation of entanglement between
multiple qubits and an oscillator by performing rotations on these qubits
conditional on the oscillator's state.

\textbf{Acknowledgments: }We thank D. J. Wineland for fruitful discussions.
This work was supported by the Major State Basic Research Development
Program of China under Grant Nos. 2012CB921601 and 2014CB921200, and the
National Natural Science Foundations of China under Grant Nos. 11374054,
11222437, and 11374083. The experiment was performed at Zhejiang University.



\renewcommand{\thefigure}{S\arabic{figure}}
\setcounter{figure}{0}
\renewcommand\figurename{Figure}
\section{Supplementary information for 'Quantum delayed-choice
experiment with a beam splitter in a quantum superposition'}

\section{Quantum circuit and qubit-resonator interaction control}
\setcounter{section}{0}
 The quantum circuit used to
implement our quantum delayed-choice experiment consists of two
phase qubits coupled to a superconducting coplanar waveguide
resonator, as shown in Fig. S.1. The sample, as described in Ref. 1,
involves four qubits, two of which are not used in our experiment
and not shown in the circuit diagram. The resonator has a fixed
frequency of $6.205$ GHz, while the frequency of each qubit is
tunable through a flux bias coil. Neglecting the effects of higher
levels of the qubits, the qubit-resonator interaction Hamiltonian in
the interaction picture is given by
\begin{equation*}
H=\hbar \sum_{j=1}^2\Omega _j\left( e^{i\Delta _jt}\sigma
_j^{+}a+e^{-i\Delta _jt}\sigma _j^{-}a^{\dagger }\right),(S1)
\label{S1}
\end{equation*}
where $\sigma _j^{+}$ and $\sigma _j^{-}$ are the raising and
lowering operators for qubit $j$, $a$ and $a^{\dagger }$ the
annihilation and creation operators of the field in the resonator,
and $\Omega _j$ is the
coupling strength between qubit $j$ and the resonator with the detuning $%
\Delta _j$. The Hamiltonian describes the coherent energy exchange
between the qubits and the resonator. When $\Delta _j\gg \Omega _j$
the photon transfer probability between qubit $j$ with the resonator
is negligible, implying their interaction is effectively switched
off. On the other hand, when $\Delta _j=0$ and $\Delta _k\gg \Omega
_k$ ($j,k=1,2$ and $j\neq k$) the interaction between qubit $j$ and
the resonator is described by the Jaynes-Cummings Hamiltonian. The
qubit frequency tunability allows us to freely control the
qubit-resonator interaction.

\begin{figure}[tbp]
\includegraphics[width=1\columnwidth]{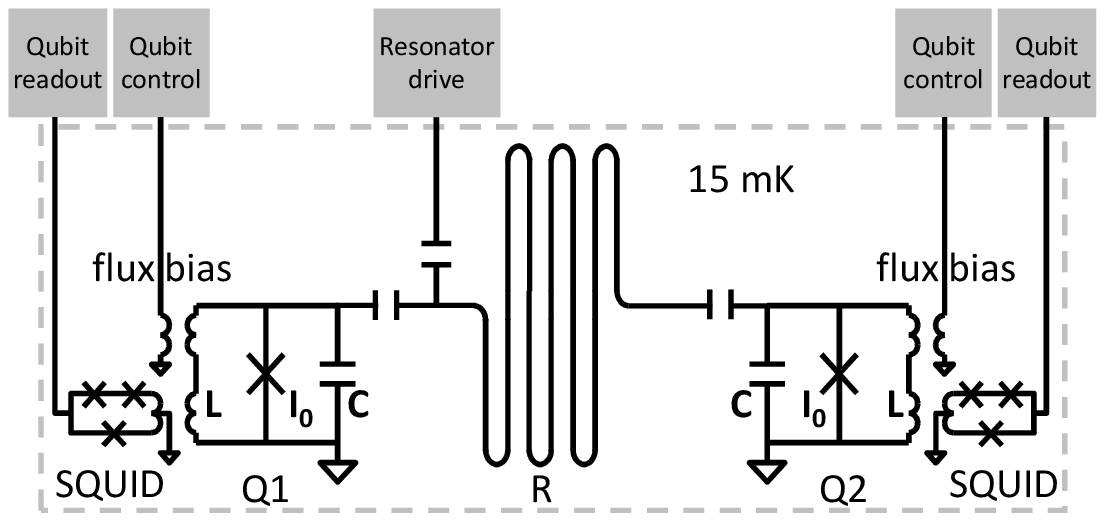}
\caption{(Color online) Circuit schematic. Two phase qubits are
coupled to a superconducting coplanar waveguide resonator via
capacitors. The detuning between each qubit and the resonator is
adjusted through a flux bias coil, enabling the relevant
qubit-resonator interaction to be effectively switched on and off.
External microwave pulses are coupled to the qubits through the flux
bias coil, and to the resonator through the capacitor on the left
side of the resonator. The coupling
strengths of the resonator to the two qubits are $2\pi \times 17.5$ MHz and $%
2\pi \times 17.7$ MHz, respectively. The energy relaxation times for
these two qubits are $520$ ns and $560$ ns, respectively, and the
Ramsey dephasing times for both qubits are about $150$ ns. The
energy relaxation time of the resonator is 3.0 $\mu $s without
measurable dephasing. }
\end{figure}

\section{The first Ramsey pulse with a coherent field in the
resonator}

The coherent field state $\left| \alpha \right\rangle $ stored in
the
resonator can be expressed as a superposition of photon number states: $%
\left| \alpha \right\rangle =\sum_{n=0}^\infty C_n\left|
n\right\rangle $, where $C_n=\exp \left( -\left| \alpha \right|
^2/2\right) \alpha ^n/\sqrt{n!} $ is the probability amplitude for
having $n$ photons. The Rabi oscillation frequency associated with
the photon-number state $\left| n\right\rangle $ is $\sqrt{n}\Omega
$. After an interaction time $t_\alpha =\pi /(4\left| \alpha \right|
\Omega )$ the state of the qubit-resonator system evolves to
\begin{eqnarray*}
\left| \Psi \right\rangle =\sum_{n=0}^\infty [ C_n\cos
(\sqrt{n}\Omega t_\alpha )| g\rangle -iC_{n+1}\sin (\sqrt{n+1}\Omega
t_\alpha ) | e\rangle ] |n\rangle.\\ (S2) \label{S2}
\end{eqnarray*}
The coherent field has a Poissonian photon-number distribution, with
the
mean photon number $\stackrel{-}{n}=\left| \alpha \right| ^2$ and variance $%
\Delta n=\left| \alpha \right| $. When the field amplitude is large,
$\Delta
n$ is much smaller than $\stackrel{-}{n}$. In this case, we have $%
C_{n+1}/C_n=\alpha /\sqrt{n+1}\simeq e^{i\vartheta }$ and
$\sqrt{n}\Omega
t_\alpha \simeq \left| \alpha \right| \Omega t_\alpha =\pi /4$, where $%
\vartheta $ is the argument of the complex amplitude $\alpha $. For $%
\vartheta =0$, the total state $\left| \Psi \right\rangle $ is
approximately a product of the qubit state $(\left| g\right\rangle
-i\left| e\right\rangle )/\sqrt{2}$ with the field state $\left|
\alpha \right\rangle $.

\section{Experimental sequence for Ramsey interference}

\begin{figure}[tbp]
\includegraphics[width=1\columnwidth]{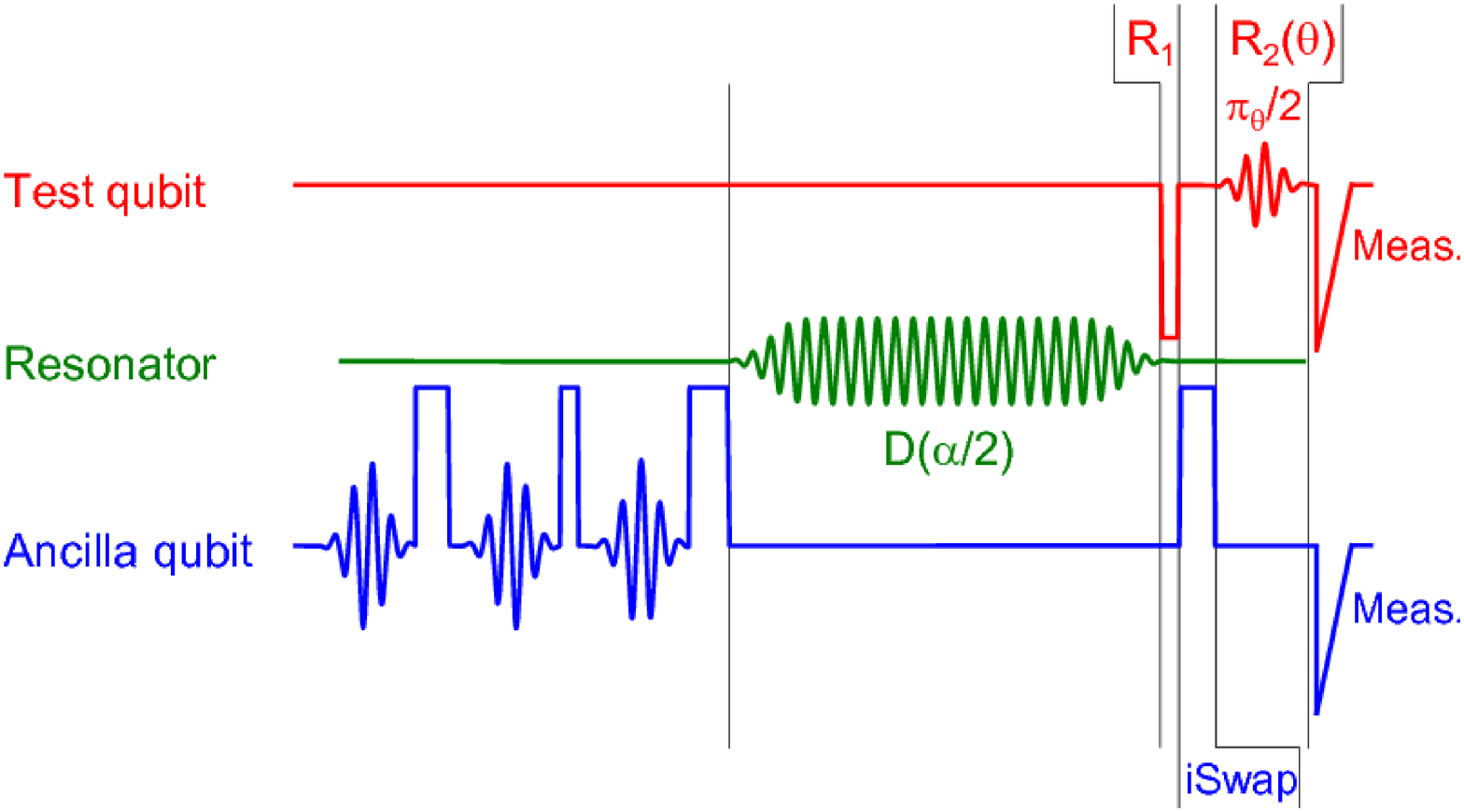}
\caption{(Color online) Experimental sequence for Ramsey
interference. Three microwave pulses (blue sinusoids) and three
qubit-resonator swap pulses (blue squares) applied alternately, are
used to generate the superposition state $\mathcal{N}(\cos \varphi
\left| \alpha \right\rangle -\sin \varphi \left| 0\right\rangle )$
with $\alpha =2$ (the cutoff is $n=4 $) in the resonator. The
ancilla qubit returns to its ground state and no further operation
is performed on it for measurement of the Ramsey signal of Fig. 2a
of the main text. The displacement pulse D($\alpha /2$) further
turns the resonator field state into $\left| \psi
_{b,i}\right\rangle $ (see Eq. (3) of the main text). R$_1$ on the
test qubit is implemented by turning on the
qubit-resonator interaction for a time t$_\alpha $ (red square), while R$%
_2(\theta ) $ achieved by applying a phase-tunable, on-resonance
microwave pulse to the test qubit (red sinusoid). For different
$\varphi$, the
resulting excited-state probability of the test qubit as a function of $%
\theta $ is measured (red triangle pulse), which forms the Ramsey
interference pattern, as shown in Fig. 2a of the main text. To
distinguish between the wave and particle outcomes as shown in Fig.
2b of the main text, after R$_1$ the state of the resonator is
examined by the ancilla qubit using an iSwap gate (the fourth blue
square), followed by ancilla state detection (the blue triangle
pulse). } \label{Fig.S2}
\end{figure}

The state $\left| \psi _{b,i}\right\rangle $ of Eq. (3) of the main
text with $\varphi \neq 0$, $\pi /2$ is generated by coherently
pumping photons into the resonator one by one through the ancilla
qubit, emploiting an algorithm theoretically proposed in Ref. 2.
Experimental implementation of this algorithm in a superconducting
resonator involves alternative, well-controlled qubit drive
operations and qubit-resonator swap operations, as detailedly
described in Refs. 3 and 4 and illustrated in Fig. S2. Here the
qubit drive operation is achieved by applying a resonant microwave
pulse through the flux bias coil. To decrease the reasonable cutoff
in the
Fock-state expansion, we first generate the supersotion state $\mathcal{N}%
(\cos \varphi \left| \alpha /2\right\rangle -\sin \varphi \left|
-\alpha /2\right\rangle )$ with $\alpha =2$ (the cutoff is $n=4$),
and then displace it in phase space by an amount $\alpha /2$,
achieving the state $\left| \psi _{b,i}\right\rangle $. The
displacement operation is performed using a microwave pulse
capacitively coupled to the resonator. For $\varphi =0$, we directly
generate $\left| \psi _{b,i}\right\rangle $ from the vacuum state by
performing the displacement operation $D(\alpha )$. We produce the
states
$\left| \psi _{b,i}\right\rangle $ for $\varphi =0$, $\pi /8$, $\pi /4$, $%
3\pi /8$, $\pi /2$, with the fidelities being $0.898\pm 0.023$,
$0.702\pm 0.021$, $0.726\pm 0.028$, $0.760\pm 0.017$, and $0.992\pm
0.004$, respectively.

After preparation of $\left| \psi _{b,i}\right\rangle $, we achieve
the first pulse R$_1$ by turning on the interaction between the test
qubit and the resonator for a time t$_\alpha $. For the observation
of the Ramsey signal displayed in Fig. 2a of the main text, R$_1$ is
directly followed by a phase-tunable, on-resonance microwave$\pi /2$
pulse R$_2(\theta )$ applied to the qubit. For different $\varphi $,
the Ramsey interference pattern is constructed by measuring the
resulting excited-state probability of the test qubit as a function
of $\theta $. To obtain the result shown in Fig. 2b of the main
text, after R$_1$ an aditional iSwap gate is applied to the
resonator field and the ancilla, as shown in Fig. S.2. The wave and
particle behaviors of the test qubit is post-selected by detecting
the state of the ancilla.

\section{Measurement of Wigner functions}

\begin{figure}[tbp]
\includegraphics[width=1\columnwidth]{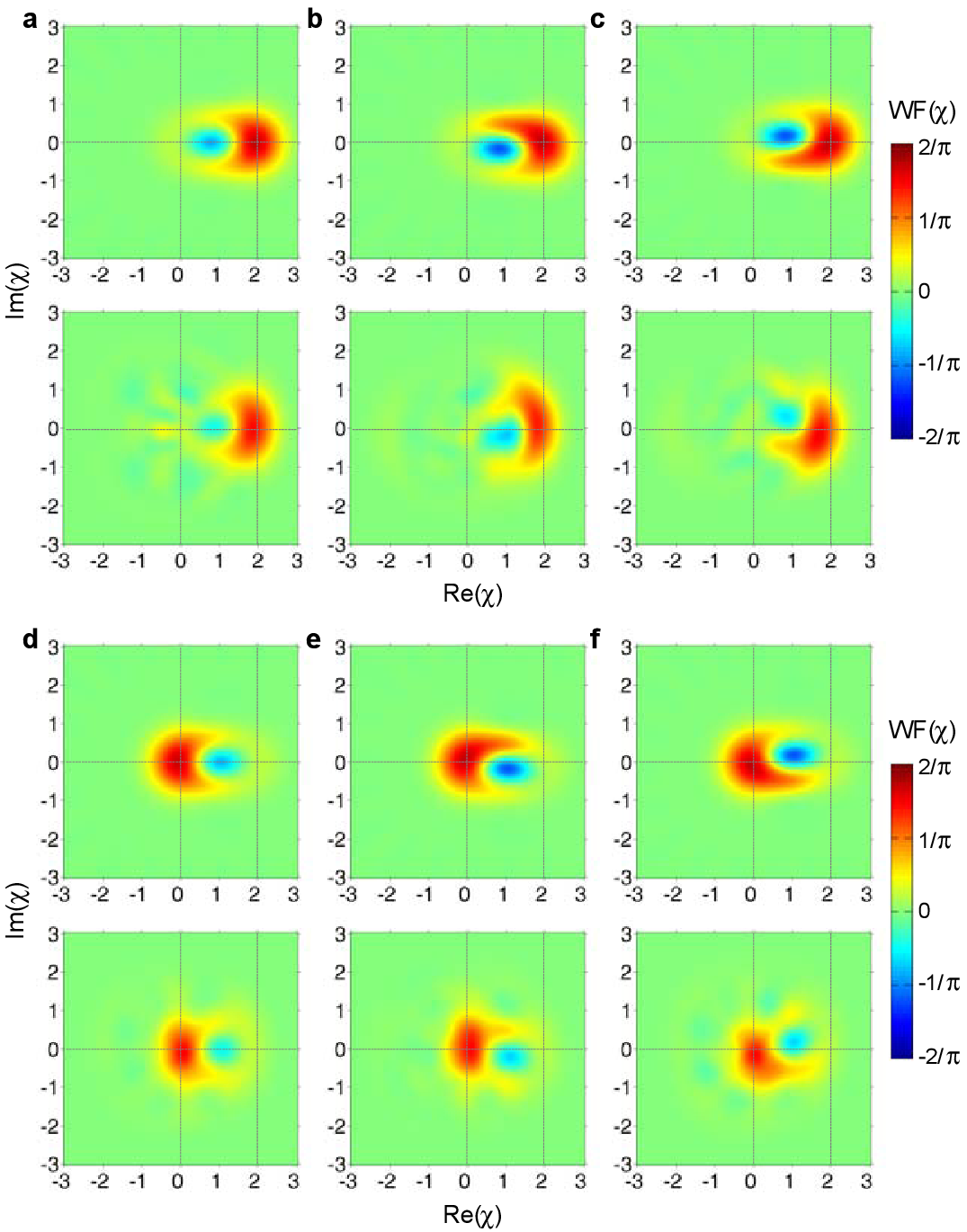}
\caption{(Color online) Wigner tomography of the quantum beam
splitter state. Panels a to c (d to f) display the WFs for $\varphi
=\pi /8$ ($3\pi /8 $). The parameters $\alpha $ and $\theta $ are as
in Fig. 3 of the main text. Shown here are the reconstructed WFs of
the resonator field for the three cases from left to right: (a and
d) The qubit state is not read-out; (b and e) The qubit is measured
in $\left| g\right\rangle $; (c and f) The qubit is measured in
$\left| e\right\rangle $. In each panel, the simulated and measured
WFs are shown in the upper and lower rows, respectively.
Experimental imperfections are not included in numerical simulation.
The minimum values of the measured WFs for $\varphi =\pi /8$ are
$-0.138\pm 0.020 $, $-0.227\pm 0.021$, and $-0.193\pm 0.024$, while
those for $\varphi =3\pi /8$ are $-0.152\pm 0.025$, $-0.218\pm
0.029$, and $-0.215\pm 0.031$, respectively. } \label{Fig.S3}
\end{figure}

To reconstruct the WF of the reduced density operator of the
resonator field without reading-out the state of the test qubit, we
perform the displacement operation $D(-\chi )$ after the second
Ramsey pulse, and then let the ancilla qubit initially in the ground
state interact with the resonator field for a variable time $\tau $,
followed by the measurement of the state of the ancilla. The
measured probability $P_e(\tau )$ for the ancilla being in the
excited state, as a function of $\tau $, is used to infer the
diagonal elements of the displaced density matrix of the resonator
field and hence the value of the WF at point $\chi $ in phase space
[$3,4$].

To map out the WF of the resonator field associated with the states
$\left| g\right\rangle $ and $\left| e\right\rangle $ of the test
qubit, we perform the joint qubit-resonator tomography, which
requires reading-out both the test and ancilla qubits
simultaneously, as described in Ref. 5.

In Fig. S3, we present the measured WFs of the field state in the
resonator for $\varphi =\pi /8$, $3\pi /8$. Panel a (d) shows the
WFs when the qubit state is traced out, while b (e) and c (f)
exhibit the WFs associated with
the outcomes $\left| g\right\rangle $ and $\left| e\right\rangle $ for $%
\varphi =\pi /8$ ($3\pi /8$), respectively. In each panel, the
simulated and measured WFs are shown in the upper and lower rows,
respectively. Experimental imperfections are not included in
numerical simulation. As expected, due to the quantum coherence
between the present and absent states of the QBS the WF for each
case exhibits a nonclassical feature around $\chi =1$. When the
qubit state is measured, the quantum interference and hence the
negativity of the WF are enhanced. These results, together with
those shown in Fig. 2 of the main text, reveal the quantum nature of
the QBS for a wide range of the parameter $\varphi $.

\section{Observing transition from quantum to classical beam splitter}

Another benefit of this experimental implementation is that it
allows the observation of the transition from a quantum to a
classical beam splitter. To demonstrate this, we now delay the
interaction between the test qubit and the resonator for a time $T$
after the QBS has been prepared in the cat state $\left| \psi
_{b,i}\right\rangle $. Then, just before the test qubit-resonator
interaction, the field density operator is given by $[6]$ \emph{\ }
\begin{eqnarray*}
\rho _b &=&\mathcal{N}^2[\cos ^2\varphi |\alpha ^{\prime }\rangle
\langle \alpha ^{\prime }|+\sin ^2\varphi |0\rangle \langle
0|\\&&-\frac 12e^{-|\alpha |^2(1-e^{-\gamma T})/2}\sin (2\varphi
)(|\alpha ^{\prime }\rangle \langle 0|+|0\rangle \langle \alpha
^{\prime }|)],(S3) \label{S3}
\end{eqnarray*}
where $\alpha ^{\prime }=\alpha e^{-\gamma T/2}$\ and $\gamma $\ is
the decay rate of a photon in the resonator; we have ignored
imperfections in the cat state preparation. In our experimental
setup the single-photon lifetime is $\tau =1/\gamma \simeq 3.0~\mu
$s. The qubit-resonator interaction time is set to $t_{\alpha
^{\prime }}=\pi /\left( 4\alpha ^{\prime }\Omega \right) $, so that
after R$_2(\theta )$ the state component $|\alpha ^{\prime }\rangle
|g\rangle $ approximately evolves to $|\alpha ^{\prime }\rangle
|\psi _w\rangle $, while $|0\rangle |g\rangle $ evolves to
$|0\rangle |\psi _p\rangle $.

\begin{figure}[tbp]
\includegraphics[width=1\columnwidth]{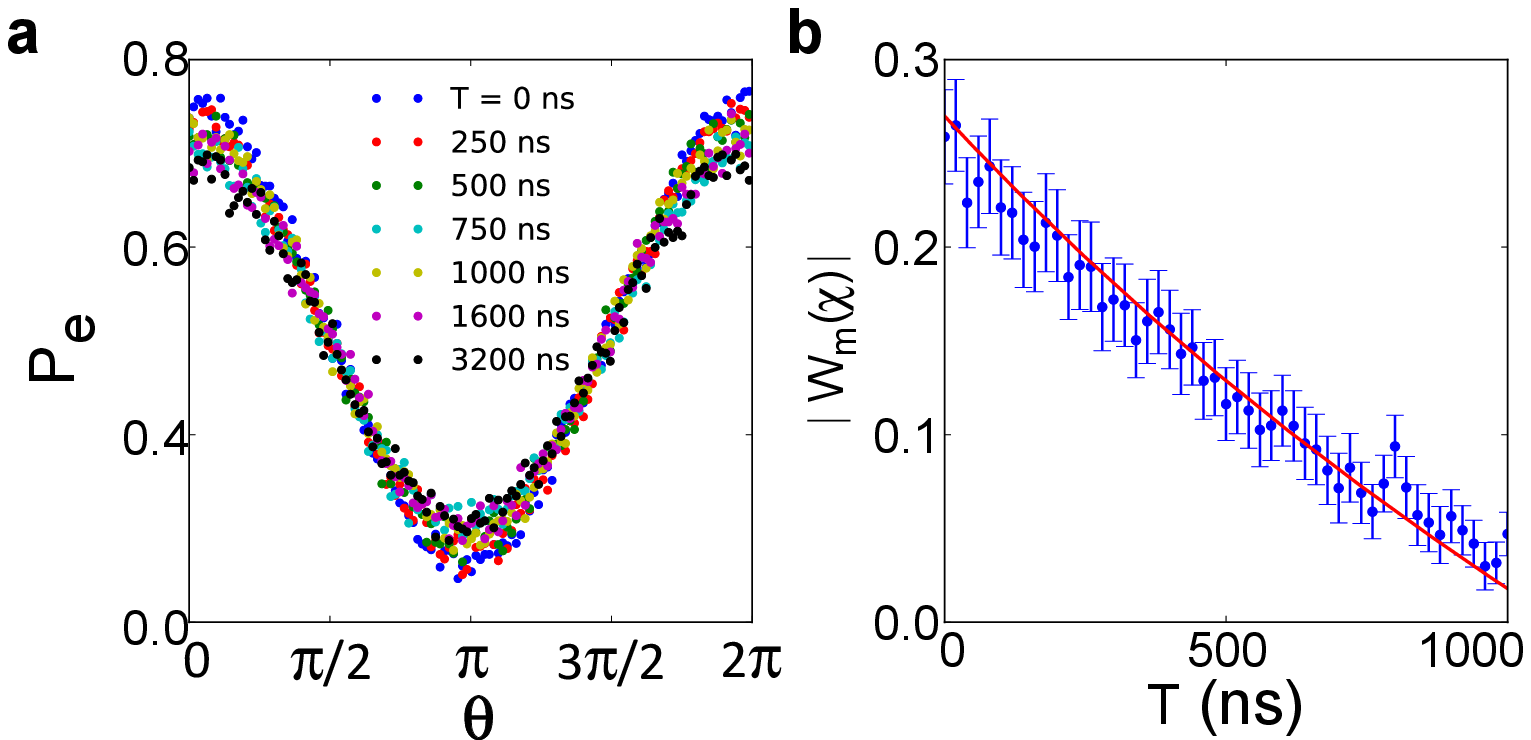}
\caption{(Color online) Transition from a quantum to a classical
beam splitter. (a) Measured Ramsey signals for different delays $T$,
defined as the interval between the end of the cat state preparation
and the start of the qubit-resonator interaction. The parameters are
the same as those in Fig.~S3. (b) Absolute value of the
negative-valued minimum of the WF of the QBS state after R$_2$,
without reading out the test qubit state, displayed versus $T$. The
damping of the quantum coherence of the QBS is characterized by the
decrease in the negativity of the WF. Error bars indicate the
statistical variance. The line is a master-equation simulation
taking the prepared cat state as the initial state for the
decoherence evolution. The resonator single-photon lifetime is
around 3.0 $\mu $s, with negligible pure dephasing.} \label{Fig.S.4}
\end{figure}

In Fig.~S4(a) we display the measured Ramsey interference signals for $%
\varphi =\pi /4$ with different delays $T$, showing that the fringe
contrast is insensitive to the field decay, as expected. However,
the quantum coherence between the active and inactive states of the
QBS degrades much faster due to decoherence. Figure~S4(b) shows the
negative-valued minimum
value of the WF as a function of delay $T$, with the WF measured after R$%
_2(\theta )$ but without reading out the test qubit state. The blue
symbols are the measured results, whereas the red curve represents
the simulated
decay, taking the prepared cat state as the initial state. For $\gamma T=1/3$%
, the amplitude of the coherent field is reduced by only $15\%$, but
the absolute value of the minimum of the WF decays almost to zero,
revealing the quick damping of the quantum coherence, which can be
defined as the sum of the off-diagonal elements of the third term of
Eq. (S3) in the photon-number\ representation $[7]$. Equation (S3)
predicts that the quantum coherence is shrunk by a factor of about
0.51 after this delay. With a cat
state of larger size and higher fidelity, as has been achieved in $[3,4,8,9]$%
, one could observe further decay of the quantum coherence of the
QBS.

\end{document}